\documentclass{article}

\usepackage{arxiv}

\usepackage[utf8]{inputenc} 
\usepackage[T1]{fontenc}    
\usepackage{hyperref}       
\usepackage{url}            
\usepackage{booktabs}       
\usepackage{amsfonts}       
\usepackage{nicefrac}       
\usepackage{microtype}      
\usepackage{lipsum}		
\usepackage{graphicx}
\usepackage{natbib}
\usepackage{doi}

\title{The Second Joint Workshop on Cross Reality}

\author{ \href{https://orcid.org/0000-0000-0000-0000}{\includegraphics[scale=0.06]{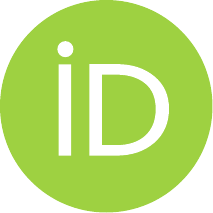}\hspace{1mm}Nanjia Wang}\thanks{Use footnote for providing further
		information about author (webpage, alternative
		address)---\emph{not} for acknowledging funding agencies.} \\
	University of Calgary\\
	Calgary, Canada\\
	\texttt{nanjia.wang1@ucalgary.ca} \\
	\And
	\href{https://orcid.org/0000-0003-3728-218X}{\includegraphics[scale=0.06]{orcid.pdf}\hspace{1mm}Yue Li } \\
	Xi'an Jiaotong Liverpool University\\
	Suzhou, China\\
	\texttt{yue.li@xjtlu.edu.cn} \\
 \And
	\href{https://orcid.org/0000-0003-2987-7634}{\includegraphics[scale=0.06]{orcid.pdf}\hspace{1mm}Francesco Chiossi } \\
	LMU Munich\\
	Munich, Germany\\
	\texttt{francesco.chiossi@um.ifi.lmu.de} \\ 
 \And
	\href{https://orcid.org/0000-0003-2987-7634}{\includegraphics[scale=0.06]{orcid.pdf}\hspace{1mm}Fabian Pointecker  } \\
	University of Applied Sciences Upper Austria\\
	Wels, Austria\\
	\texttt{fabian.pointecker@fh-hagenberg.at} \\ 
 \And
	\href{https://orcid.org/0000-0001-6181-1673}{\includegraphics[scale=0.06]{orcid.pdf}\hspace{1mm}Lixiang Zhao   } \\
	Xi'an Jiaotong-Liverpool University\\
	Suzhou, China\\
	\texttt{lixiang.zhao17@student.xjtlu.edu.cn} \\  
 \And
	\href{https://orcid.org/0000-0001-6181-1673}{\includegraphics[scale=0.06]{orcid.pdf}\hspace{1mm}Daniel Zielasko   } \\
	Trier University\\
	Trier, Germany\\
	\texttt{zielasko@uni-trier.de} \\
}

\date{}

\renewcommand{\headeright}{JWCR'24}
\renewcommand{\shorttitle}{JWCR'24}

\hypersetup{
pdftitle={Joint Workshop on Cross Reality},
pdfsubject={q-bio.NC, q-bio.QM},
pdfauthor={Francesco Chiossi},
pdfkeywords={Cross-Reality, Mixed Reality, Virtual Reality, Augmented Reality, Augmented Virtuality, Spatial Computing},
}

\begin{document}
\maketitle

\begin{abstract}
The 2nd Joint Workshop on Cross Reality (JWCR'24), organized as part of ISMAR 2024, seeks to explore the burgeoning field of Cross Reality (CR), which encompasses the seamless integration and transition between various points on the reality-virtuality continuum (RVC) such as Virtual Reality (VR), Augmented Virtuality (AV), and Augmented Reality (AR). This hybrid workshop aims to build upon the foundation laid by the inaugural JWCR at ISMAR 2023, which successfully unified diverse CR research communities.
The workshop will address key themes including CR visualization, interaction, user behavior, design, development, engineering, and collaboration. CR Visualization focuses on creating and displaying spatial data across the RVC, enabling users to navigate and interpret information fluidly. CR Interaction delves into natural user engagements using gestures, voice commands, and other advanced techniques to enhance immersion. The study of CR User Behavior and Experience investigates how users perceive and interact within these hybrid environments. Furthermore, CR Design and Development emphasizes creating effective CR applications using innovative processes and tools, while CR Collaboration examines methods for fostering teamwork in mixed reality settings.
\end{abstract}


\section{Introduction}
Cross Reality (CR) \cite{auda2024scoping} is an emerging technology that focuses on the concurrent usage of or the transition between multiple systems at different points on the reality-virtuality continuum (RVC), including Virtual Reality (VR), Augmented Virtuality (AV), and Augmented Reality (AR). CR has gained significant attention in recent years due to its potential for revolutionizing various research and industry areas where users need to comprehend and explore spatial data and its relevant information in different forms \cite{chiossi2022it}. It is expected that in the near future, more CR applications will arise to allow users to transition along the individual stages of the RVC or to collaborate in-between these stages to use their distinct advantages and mitigate their potential problems.     

\subsection{Themes and Topics of Interest}
Last year, the 1st Joint Workshop on Cross Reality \cite{Liang2023joint} held in conjunction with the IEEE ISMAR 2023 brought together the community of CR researchers from five separate workshops on individual CR topics. The workshop drew a lot of attention where participants established the research area further. There were over 50 participants joined the 1st JWCR in person with over 20 participants joined online.    This is a follow up workshop to explore the new challenges, solutions, and opportunities in the field of CR and provide a comprehensive overview of the related research on the design of interactive techniques for effective CR visualization, interaction, collaboration, user experience, behavior and engineering. We are particularly interested in blurring the lines between and merging the virtual and physical worlds to support users in understanding data in different forms. The workshop provides a forum for researchers from VR/AV/AR/MR, visualization, HCI, and related fields to present their technical and systems papers that introduce new approaches, ideas, discussions, and applications.

\section*{Topics of Interest}

The field of Cross Reality (CR) is a rapidly evolving area of research that encompasses various aspects of technology and human interaction. CR integrates and transitions between multiple systems on the reality-virtuality continuum, including Virtual Reality (VR), Augmented Reality (AR), and Augmented Virtuality (AV). This convergence of real and virtual worlds opens up numerous possibilities for innovation in how we perceive, interact with, and utilize digital information in our daily lives and professional activities.

CR research covers a broad spectrum of topics, aiming to enhance user experiences and functionalities across different reality-virtuality interfaces. The integration of virtual and physical environments presents unique challenges and opportunities in design, interaction, and application development. Researchers and practitioners are exploring new methodologies and technologies to create more immersive, interactive, and effective CR systems.

Key areas of interest include, but are not limited to, the following:

\begin{itemize}
  \item Cross Reality environment design: virtuality and reality
  \item Computational and Adaptive Cross Reality Systems
  \item Visual representations in cross-reality systems
  \item Cross Reality transitions across multiple interfaces
  \item Cross Reality interaction
  \item Multimodal interaction, perception, and cognition
  \item Real-world tool use and tangibles as input to CR systems
  \item Collaborative Cross Reality immersive analytics
  \item Collaboration across the reality-virtuality continuum
  \item Cross Reality user experience
  \item Cross Reality user behavior
  \item Cross Reality tools, frameworks, and APIs
  \item Testing Cross Reality systems
  \item Design guidelines for Cross Reality applications
  \item Evaluation of Cross Reality experiences and systems
\end{itemize}

Researchers in these areas are encouraged to contribute their insights and findings to advance the state of CR technology and its applications. By addressing these topics, we aim to foster a deeper understanding of CR and its potential to transform various domains, including education, healthcare, industry, and entertainment.

\subsubsection{CR Visualization}
CR Visualization refers to the ability to create and display abstract or inherently spatial data along the RVC. The visual representations of a Cross-Reality system offer different degrees of virtuality to users and enable users to seamlessly move back and forth across the reality-virtuality continuum.

\subsubsection{CR Interaction}
CR Interaction allows users to naturally and intuitively engage with real and virtual objects and environments. CR enables various forms of interaction, including hand gestures, voice commands, haptic feedback, and eye-tracking. Advanced interaction techniques for object manipulation, locomotion, and spatial interaction enhance the level of immersion and interactivity in CR experiences, enabling users to seamlessly perform complex tasks and interactions. 

\subsubsection{CR User Behavior and Experience }
CR User Behavior and Experience are concerned with how users perceive, process andrespond to visual information in CR environments. CR technology enables users to immerse themselves in virtual or augmented reality spaces, allowing them to interact with and seamlessly transition between virtual and physical environments and objects.

\subsubsection{CR Design, Development, and Engineering }
CR Design, Development, and Engineering are concerned with processes, methods, frameworks, and tools that increase the effectiveness and efficiency of design and development teams tasked with creating CR applications and systems. Testing and evaluation are a substantial part of these processes. Additionally, there are still many open challenges in enabling end users, such as developers, domain experts, and users with minimal programming experience, to craft, program, and tailor CR applications and experiences according to their goals, interests, needs, and abilities. There is a growing interest in utilizing authoring tools powered by Generative AI for prototyping CR applications.

\subsubsection{CR Collaboration}
CR Collaboration is an area of growing interest, as it allows users to work together in real and virtual environments using different interaction techniques and interfaces with different degrees of virtuality. Collaborative CR experiences have the potential to enhance local, remote, and hybrid communication, foster creativity, and improve decision-making in various domains such as design, training, and education.

\subsubsection{CR Applications }
CR Applications have the potential to enhance the effectiveness and efficiency of real-life tasks in diverse industries. However, there needs to be more empirical evidence to support claims made about the advantages of serious CR; namely, more CR applications need to demonstrate their benefit in solving real-life tasks and be evaluated by empirical studies with the respective user group. 

\subsection{Workshop Format}
This will be a half-day to a full-day workshop with a mixture of invited keynote speeches, presentations of short position papers, and panel discussion on topics derived from the position papers. Position papers will each be presented for ten minutes, while thirty to sixty minutes will be allocated for the group discussion activity.   It will be a hybrid event supporting both face-to-face and online participation. A discord channel will be created, enabling people to discuss the workshop topic both before and after the ISMAR event.

\subsection{Workshop Contributions}
We welcome position paper submissions from 2-4 pages long, excluding references. All paper submissions must be in English. Paper quality versus length will be assessed according to a contribution-per-page judgment. All submissions will be accepted or rejected as workshop papers. All accepted papers will be archived in the IEEE Xplore digital library. Detailed submission and review guidelines are available on the workshop website: \url{https://cross-realities.org/2024/}

\subsection{Audience}
Based on the statistics of the 1st JWCR at ISMAR 2023, we are expecting 15 to 20 submissions with paper presentations at the workshop. An additional 20 to 30 participants will join the panel discussion with presenters. 

\subsection{Important Dates}

We encourage all interested researchers and practitioners to note the following important dates for the 2nd Joint Workshop on Cross Reality (JWCR'24):

\begin{itemize}
  \item \textbf{July 26, 2024:} Papers submission deadline
  \item \textbf{August 9, 2024:} Notification of acceptance
  \item \textbf{August 26, 2024:} Camera-ready submission deadline
  \item \textbf{October 25, 2024:} Workshop date
\end{itemize}

\subsection{Call for Participation}

We are pleased to invite you to the 2nd Joint Workshop on Cross Reality (JWCR'24), held in conjunction with ISMAR 2024. Cross Reality (CR) is an emerging technology that enables concurrent usage and transitions between multiple systems across the reality-virtuality continuum (RVC), including Virtual Reality (VR), Augmented Virtuality (AV), and Augmented Reality (AR). This technology has garnered significant attention for its potential to revolutionize research and industry by allowing users to explore and comprehend spatial data in diverse forms.

Following the success of the inaugural JWCR at ISMAR 2023, which united the CR research community from five separate workshops, this year's workshop aims to further explore new challenges, solutions, and opportunities in the field. We seek to provide a comprehensive overview of research on interactive techniques for effective CR visualization, interaction, collaboration, user experience, behavior, and engineering. Our goal is to blur the lines between the virtual and physical worlds, enhancing users' ability to understand and manipulate data across different forms.

\subsection{Organizers}
\paragraph{Nanjia Wang}is a Ph.D. student at the SEER lab (\url{https://seriousxr.ca/}) under the supervision of Dr. Frank Maurer. His work focuses on single-user cross-reality (CR) applications \cite{wang2022design}. His research goal is to exploit the potential benefits of single-user CR applications, including the ability to transition between spaces on the Milgram’s Reality Virtuality Continuum (RVC) that are most suitable for specific tasks \cite{aigner2023cardiac, wang2020individual}. Nanjia is a member of the 1st Joint Workshop on Cross Reality program committee.

\paragraph{Yue Li}received her PhD degree from the University of Nottingham in 2020 with the Postgraduate Award. She is an Assistant Professor at Xi’an Jiaotong-Liverpool University (\url{https://imyueli.github.io/}). Her research interest is in the field of HCI, with particular emphasis on Cross Reality technologies and their use in cultural heritage and education \cite{ch2020effects,xu2024cubemuseum, he2024data}.

\paragraph{Francesco Chiossi} (\url{www.francesco-chiossi-hci.com} is a PhD researcher at LMU Munich with a background in applied cognitive science. He focuses on implicit measures of user behavior and context, such as electrodermal activity \cite{chiossi2023adapting} and EEG \cite{chiossi2023designing}, as an implicit input to design and evaluate physiologically adaptive systems across the reality-virtuality continuum \cite{chiossi2024impact}. 

\paragraph{Fabian Pointecker} received his Master’s degree in Human-Centered Computing from the University of Applied Sciences Upper Austria and is a member of the program committee of the 1st Joint Workshop on Cross Reality. His research interests are visualization and transition techniques in the area of CR (\url{https://hive.fh-hagenberg.at/}). He is currently focusing on visual transition techniques that allow the user to seamlessly move along Milgrams’s Reality Virtuality Continuum.

\paragraph{Lixiang Zhao}is a Ph.D. student at VACT lab in Xi'an Jiaotong Liverpool University (\url{https://lixiangzhao98.github.io/}), supervised by Dr. Lingyun Yu. His research focuses on Scientific Visualization (SciVis) and HCI. His work is dedicated to design and evaluate the spatial interaction techniques \cite{zhao2023metacast} and visualization techniques \cite{li2023immerview} within CR and VR environments, tailored to specific scientific domains including astronomy and biomedicine.

\paragraph{Daniel Zielasko} is currently a postdoctoral researcher at the HCI Group at the University of Trier. He received his Ph.D. in 2020 at the Virtual Reality and Immersive Visualization group at RWTH Aachen University for studying desk-based Virtual Reality. He received his Master's degree in Computer Science in 2013 at RWTH Aachen University.

He focuses on the integration of VR technologies into existing professional workflows and collaborates with neuroscientists, archaeologists, and psychologists in various interdisciplinary projects such as the EU Flagship Project, "The Human Brain Project (HBP)," and the SMHB (Supercomputing and Modeling for the Human Brain). His special interest is the prevention of cybersickness \cite{zielasko2024progression} and the development of convincing and innovative 3D user interfaces \cite{wang2024user, zielasko2017remain}.

\bibliographystyle{unsrtnat}
\bibliography{references}  






\end{document}